\documentclass{article}
\usepackage{amssymb}
\usepackage{graphicx}

\input{tcilatex}

\begin{document}

\title{Models, Calculation and Optimization of Gas Networks, Equipment and
Contracts for Design, Operation, Booking and Accounting}
\author{L. A. Ostromuhov \thanks{\textit{Dr. L. A. Ostromuhov, Wingas
Transport GmbH, Baumbachstr. 1, 34119 Kassel, Germany. E-mail:
leonid.ostromuhov@wingas-transport.de}} \thanks{%
The paper is presented on the World Symposium on Computing in the Gas
Industry, 26-28.04.1999 in Florence, Italy. }}
\date{}
\maketitle

\begin{abstract}
\qquad There are proposed models of contracts, technological equipment and
gas networks and methods of their optimization. The flow in network
undergoes restrictions of contracts and equipment to be operated. The values
of sources and sinks are provided by contracts. The contract models
represent (sub-) networks. The simplest contracts represent either nodes or
edges. Equipment is modeled by edges. More sophisticated equipment is
represented by sub-networks. Examples of such equipment are multi-poles and
compressor stations with many entries and exits. The edges can be of
different types corresponding to equipment and contracts. On such edges,
there are given systems of equation and inequalities simulating the
contracts and equipment. On this base, the methods proposed that allow:
calculation and control of contract values for booking on future days and
for accounting of sales and purchases; simulation and optimization of design
and of operation of gas networks. These models and methods are implemented
in software systems ACCORD and Graphicord as well as in the distributed
control system GAMOS used by Wingas, Germany. As numerical example, the
industrial computations are presented.
\end{abstract}

\tableofcontents

\section{Introduction}

In the paper, the contract models are proposed, which represent (sub-)
networks. These contract models are used for booking, accounting, estimation
of contract values, for control and operation of contracts, and for network
optimization.

In addition, equipment is modeled by edges and sub-networks. The edges can
be of different types. A type of an edge correspond to a type of equipment
and contracts. On such edges, there are given systems of equation and
inequalities simulating the operation of equipment and contracts. In case of
a gas network, an edge might have a type of a pipe, compressor, compressor
station, control valve, shut-off valve, and so on.

The models of contracts and equipment construct a base for methods for
balancing and optimization both of contracts and of gas networks for design,
planning, control, operation, booking and accounting. \ 

These models, methods and functions describing contracts and equipment are
realized in the software systems ACCORD and GRAPHICORD.

The proposed here contract models are implemented in the distributed control
system GAMOS used by Wingas, Germany. The program implementation in GAMOS is
provided by Dr. Kaemmerling.

\bigskip

The software ACCORD has been developed as application of economic and
operational steady state simulation and optimization of gas pipeline
networks.

Any network topology and in principle almost any objectives are available.
Minimization of fuel gas and purchased electricity, minimization of cost of
purchased gas, injection and withdrawal optimization for storage facilities
within the network, profit maximization with cost consideration are provided
for any network. Both the quality and supply tracking are available. It
makes possible to produce the cost tracking, to estimate and to establish
the optimal selling prices. Compressors can be considered both with and
without description of their operating ranges or characteristics.

ACCORD has graphical user interface and interface to the third party
programs e.g. to the distributed control systems. The programs providing
these types of interface construct a family named GRAPHICORD Suite. The
software GRAPHICORD provides Graphical user interface; database support;
interface with Geographic information system; interface to Distributed
control systems as well as to the third party systems; call of the third
party programs; flie management; scenario management; means to the search of
information in the networks saved; etc. The interface to the third party
programs has input and output in the form of text files. It allows to
connect ACCORD with a graphic editor; with geographic information system;
with Distributed control systems; with spreadsheet such as MS Excel. These
connections are used to create a network; to receive initial values of
supplies, demands, and gas parameters; to produce and present a graphic
output. GRAPHICORD supports such databases as MS Access, Microsoft SQL
Server, Oracle. Selection of the type of database is provided by the user.

Steady state optimization by ACCORD had been connected with programs that
perform a dynamic simulation of gas networks.

ACCORD is integrated in the distributed control system such as GAMOS that
consists of SCADA and various high level functions.

As numerical examples, the industrial computations are presented.

\bigskip

ACCORD is an abbreviation for Algorithmic and software Complex of
Constructive nonlinear Optimization with Restriction on Discrete-Continuous
network variables. It is intended for economic and operational steady state
optimization of gas pipeline networks. There is no restriction on network
topology in ACCORD.

The general objectives of optimization include the following:

\qquad -- to plan the installation of a network;

\qquad -- to optimize the route plan of pipelines and location of equipment;

\qquad -- to select pipe diameters;

\qquad -- to specify facilities;

\qquad -- to minimize the expenditure;

\qquad -- to select supplies and demands;

\qquad -- to optimize network operation, etc.

There are available in ACCORD such objective functions as:

\qquad -- operational expenditure minimization;

\qquad -- network flow maximization;

\qquad -- set-point deviation minimization;

\qquad -- profit maximization;

\qquad -- minimization of weighted average cost of gas;

\qquad -- specific transport cost minimization;

\qquad -- injection and withdrawal optimization for storage facilities in
pipeline network, etc.

There are available such searching variables as:

\qquad -- pressure ratio of compressor stations;

\qquad -- output pressure of control valves;

\qquad -- equipment of compressor stations;

\qquad -- flow values of user-chosen supplies and demands;

\qquad -- pressure in a user-chosen node.

A network state is a distribution of flow and pressure. The feasible network
state is received by ACCORD automatically as a result of optimization
problem.

Both the quality and supply tracking are available. It makes possible to
produce the cost tracking and to establish the optimal selling prices.

Compressors can be considered in ACCORD both with and without description of
their operating ranges and characteristics of macnines.

By computation of pressure drop, the modelling procedures are used which
simulate the operation of technical equipment. Therefore computation of
pressure drop can be sophisticated.

The problems solved in ACCORD are to find the optimal set-points for
supplies, demands, and pressures and to make the optimal choice of equipment
on compressor stations to optimize both operational transport costs and
costs for purchased gas in order to optimize profit. In this class of
problems the objective functions depend on both flow and pressure. Minimal
cost and maximal flow problems are generalized. The typical problem consists
in the optimal choice of:

\qquad -- dependencies between flow and pressure from the given families;

\qquad -- node intensities, i.e. values of supplies and demands;

\qquad -- flow;

\qquad -- pressure.

This is a class of problems belonging to the area of large scale nonlinear
mixed integer programming or constrained discrete-continuous optimization on
general networks. The developed optimization method represents a branching
multi-level computational process. It is based on nonlinear and integer
programming and on graph theory. Its main characteristic feature consists in
obtaining of dominant solutions on the subnetworks, which represent the
specially constructed fragments of the network.

ACCORD has an input and output interface in the form of text files. It
allows to connect ACCORD with a graphic editor or geographic system to
create a network; to receive initial values of supplies, demands, and gas
parameters; to monitor the runs of optimization; to produce and present a
graphic output. Due to its interface, ACCORD can be connected with any other
simulation program and SCADA system to provide steady state optimization of
gas networks with economic, business or operational objectives.

\bigskip

The paper consists of 6 significant sections and supplemented sections
containg notations, figures, and tables.

Section 1 is an introduction while section 6 presents conclusions.

In section 2, mathematical model and problem formulation are presented.

In section 3, the developed methods of the continuous - discrete nonlinear
optimization on networks are briefly described.

Section 4 contains a numerical example as a result of industrial
computational experience. For optimal operation, a gas network was optimized
that consist of a central distributed ring with large - scale supply and
transport pipelines containing ten - th compressor stations .

Section 5 presents the quality, supply and cost tracking computed by ACCORD
in consequence of optimization. The graphic presentation of the tracking is
made on a path between two nodes in a gas network.

Section 6 contains conclusions.

\section{Models and problems}

\subsection{Analysis of models used in gas supply companies}

Let us consider a set of models and functions which are widespread in gas
transport, distribution and gas supply companies:

\qquad 1) gas demand forecast for every client;

\qquad 2) definition of set of gas suppliers and corresponding purchase
quantities;

\qquad 3) definition of gas injection and withdrawal quantities for every
gas storage;

\qquad 4) optimization of contracts for gas sale and purchase, including
both gas exchange with adjacent companies and injection or withdrawal for
different gas storage;

\qquad 5) contract optimization with restrictions arising due to gas network;

\qquad 6) gas price investigation, calculation and definition for sale and
purchase for long and short term planning;

\qquad 7) determination of sold and purchased gas quantities in order to
prepare billing;

\qquad 8) settlement and confirmation of accounts for sold and purchased gas
for every supplier and consumer and for every client station;

\qquad 9) SCADA functions both for sale and purchase and for gas network
(SCADA is abbreviation for supervision, control and data acquisition);

\qquad 10) supervision of actual contract fulfilment;

\qquad 11) optimization of gas network state;

\qquad 12) simulation and optimization of development of gas network.

The above functions allow to conclude that the following models play a
central role:

\qquad 1) contract;

\qquad 2) contract partner: supplier, consumer, gas storage, etc.;

\qquad 3) client station;

\qquad 4) gas network;

\qquad 5) flow network.

At present, these models have different development degrees.

The network modelling went through several development steps. It derives its
sources from Kirchhoff. It follows the development periods of graph theory,
flow networks, dynamic hydraulic simulation of networks, and network steady
state optimization. It should be pointed out that models for network dynamic
optimization are not developed yet so that they could be used industrially.

There are various contract models. They have a common feature that they are
not connected with a model of gas network. It could seem that contract
models are significantly simpler than network models. More detailed
consideration shows this is not the case.

A contract model source is the contract text. To originate a model, the text
must accept a formalization. Then a degree of detailed elaboration of the
formalization depends on the function wherein the model is used. Hence a
contract model depends on functions wherein the contract or contract
restrictions are used. Therefore there is a contract model family
corresponding to family of functions of planning, control, invoicing, etc.

Questions arise, how these models are connected with each other; could they
be connected formally and therefore automatically; do they accept a uniform
representation. In the paper, there is proposed a contract model which
represents an object consisting of:

\qquad -- a graph;

\qquad -- an ordered set of feasible intervals of gas parameters such as gas
demand, pressure, calorific value;

\qquad -- price calculation procedures corresponding usually to the above
intervals of gas parameters;

\qquad -- integral quantities of gas.

The paragraphs of a contract or of an operational agreement could be
represented by elements attributed to edges of the contract network, i.e. to
the connections between contract partners.

In the simplest case, the contract network consists of a node, with which
the feasible intervals of gas parameters and the procedures for calculation
of gas prices and transport expenses correspond.

Let us remark that an edge can be taken instead of node equally in that
case. Moreover, it can be taken more than one edge or node, e.g. one edge
for actual booking, the second one for communication, the third for
accounting, and the force for operation. In particularly, this approach can
be useful for the so called entry-exit economic models of contracts.

Such a contract model in form of a network causes different merits. It
enables a simple, visualizable, structured and unified representation of
contracts. It admits an automatic choice of degree of detailed elaboration
of a contract model, i.e. auto-modelling due to automation of procedures of
reduction or expansion (re-reduction) of contract network.

A comparison of planning and invoicing functions allows to find a lot of
similarities. In both cases, it handles with a gas distribution between the
sources or targets correspondingly. But direction of time as well as of
cumulative or de-cumulative operation is opposite by planning and invoicing.

It is sufficient to compare a planning function of definition of purchase
quantities covering an expected demand with an invoicing function of
distribution of gas quantities metered on a client station over the
contracts and contract conditions.

By planning function such as demand coverage, the client demands are
accumulated to define a total demand and to distribute demand through
suppliers. The delivery ways from supplier to customer could be considered
either implicitly by restrictions of maximal delivered gas quantities or
explicitly by explicit representation of gas network structure. Time is
directed forwards, in the future.

By invoicing, the metered gas quantities have to be distributed i.e.
de-cumulated on a client station in accordance with contracts, contract
partners and prices. For the most contracts, the gas flow ways from
measurement places to contract partners could be considered explicitly by
explicit representation of client station structure. The graph theory gives
a suitable means for that. Time is directed backwards since information
about the earlier purchased gas is processed.

Thus, the uniform modelling of pipelines, client stations and contracts by
means of networks allows to represent their relations and producible
operations explicitly, to visualize models, and it establishes a base for
strict mathematical problem formulation.

\subsection{Contract and client station models}

\subsubsection{Contract models for planning and control of gas networks and
for invoicing.}

The simplest representation of a contract is an object consisting of:

\qquad 1) a node;

\qquad 2) an interval of feasible demand;

\qquad 3) an interval of feasible pressure;

\qquad 4) values of demand and pressure;

\qquad 5) a procedure for payment calculation.

A more complicated contract having an ordered set of feasible intervals of
demand and pressure with corresponding procedures for invoicing shall be
represented by a (sub-) network.

Such a model causes different merits. It enables a simple, visualizable,
structural and unified representation of contracts.

The unified contract representation is used for the following purposes and
by the following implementations:

\qquad -- both for technical and economical objectives;

\qquad -- for technical and contractual supervision of gas network in
dispatcher center;

\qquad -- for short and long term planning;

\qquad -- by gas balancing for automation of checking and achievement of gas
balance;

\qquad -- by data acquisition from client stations, i.e. gas meter stations;

\qquad -- by data preparation for settlement of accounts;

\qquad -- to construct the data base of gas company.

\subsubsection{Models of client and shut-off stations.}

A client station is a gas meter station. It consists of:

\qquad -- shutting-off devices;

\qquad -- means providing measuring of gas flow, pressure, temperature, and
other;

\qquad -- valves;

\qquad -- control valves;

\qquad -- service pipes.

A shut-off station has the same structure as a client station. A difference
is only that the flow meter and control valves are used rather seldom by
shut-off stations. So the client and shut-off stations can be modelled in
the same way.

A gas meter station has a technical structure plotting as a scheme. A
configuration scheme is represented by a (sub-) network. Model of a gas
meter station shall be a (sub-) network.

Station modelling by networks causes the same merits as contract modelling
by networks.

Hence networks are used as models for gas networks, contracts, client and
shut-off stations.

\subsection{Network models}

\subsubsection{Network models for hydraulic simulation and optimization.}

A network is modelled as a graph that is equiped with flow and which nodes
and edges possess parameters. Nodes and edges possess technical parameters.
Pressure is considered explicitly. It is a nodal parameter. Restrictions are
formulated in terms of both pressure and flow.

Technical equipment as pipeline sections, compressor stations, control
valves and valves are represented by edges of the network. The complicated
compressor stations can be represented by sub-networks of the network.

Client and shut-off stations are represented by nodes of the network. The
complicated client and shut-off stations can be represented by sub-networks
of the network. If a contract is connected with only one client station and
this station is represented by a node of the network then the contract is
represented by the same node too. More complicated contracts can be
represented by sub-networks of the network.

By computation of pressure drop, the modelling procedures are used which
simulate the operating of technical equipment. Therefore computation of
pressure drop can be very sophisticated.

Exactly as in network operation, a pipeline capacity is represented
implicitly. This means that pipeline capacities are given due to maximal and
minimal pressure limits in nodes of network. Pressure limits arise owing to
technical and contractual restrictions.

There are two types of hydraulic simulation: a dynamic simulation and a
steady state simulation.

In the dynamic simulation, the partial differential equations are used to
model hydraulic. The network parameters such as flow and pressure depend on
time.

In the steady state simulation, the algebraic equations are used. The
network parameters such as flow and pressure are taken at a certain moment
or in average.

A network optimization based on the hydraulic simulation considers
restrictions and objective functions which are formulated in terms of both
pressure and flow. As a result of optimization, the feasible and optimal
hydraulic state of the whole network must be found and simulated.

A network dynamic optimization, i.e. the network optimization based on the
dynamic hydraulic simulation is not developed anywhere yet. A network steady
state optimization, i.e. the network optimization based on the steady state
hydraulic simulation is provided in Wingas by methods and software ACCORD
developed there.

\subsubsection{Using the optimization based on hydraulic simulation by
planning.}

For short term planning, the network steady state optimization by ACCORD is
a useful tool. When planning for serial time periods, the network steady
state optimization can be used. For this purpose it must run several times
sequentially for sequential moments. Practically for long term planning it
is possible rather often. But, for some cases, it can be expensive if not
prohibited due to complexity of problem.

For example, a feasibility of the network state must be checked every year
for an extremely cold winter and for a summer which follows after the
extremely cold winter. Hence a planning for e.g. 25 years requires at least
50 feasibility checking of the network state. Every feasibility checking of
the network state is reduced to the solution of a network optimization
problem based on the steady state hydraulic simulation. It means that ACCORD
must run 50 times. For complex networks represented in detail, it requires
such user efforts that cannot be completed in a very short time.

\subsection{Flow network models}

A network is modelled as a graph that is equiped with flow. Edges dispose of
flow. For nodes the value of demand or supply are considered. Simulation of
operation of technical equipment is not used. Hydraulic simulation is not
used either.

Pressure is considered implicitly. Restrictions are formulated in terms of
maximal flow that is capacity for every edge. Pipeline capacities must be
given as input data. Capacity is an idealization that is not met by network
operation. Capacities simulate technical and contractual restrictions such
as maximal and minimal pressure limits. A capacity does model implicitly the
pressure limits. Any client station and any contract is represented by a
node. If necessary, a station and a contract can be represented by a
sub-network of the network.

To optimize gas network without hydraulic simulation, flow models are use
such as in the minimal cost network flow problem.

Let $G=(V,E)$ be a network with a node set $V$ and an edge set $E$. The
minimal cost network flow problem consists in:

\begin{equation}
minimize\quad \sum_{(i,k)\in E}f_{ik}(q_{ik})
\end{equation}%
subject to:

\begin{equation}
\sum_{j}q_{ji}-\sum_{k}q_{k}=Q_{i},\quad i\in V,
\end{equation}

\begin{equation}
q_{ik}^{min}<q_{ik}<q_{ik}^{max},\quad (i,k)\in E,
\end{equation}%
where: $q_{ik}$ is flow in arc $(i,k)$; $f_{ik}(q_{ik})$ is cost of flow $%
q_{ik}$ ; $Q_{i}$ is supply or demand in node $i$ .

The minimal cost network flow problem is much easier as a network
optimization with hydraulic simulation. There are efficient methods
delivering a solution of this problem.

For long term planning for serial time periods, it is worthwhile to use a
minimal cost network flow problem as a compromise simplification. It must
run several times sequentially for sequential moments for this purpose. Due
to efficient solution methods, a planning for 25 years requiring 50 - 100
solutions of minimal cost network flow problem seems realistic.

The correct investigation of capacities can turn out as a weakest point in
such a simplified approach. Here is useful a method for precise definition
of pressure limits in a gas network. This method we have developed in \cite%
{9 0-93}. Normally by long term planning, the supplies and demands are
considered as constant in every time period. Hence it is allowed to use the
said method for sub-networks that are defined by connected subsets of nodes
which include terminal nodes and nodes of compressor stations.

\subsection{Mathematical model and problem formulation}

Let be given a connected undirected network $G=(V,E)$ with a set of nodes $V$
and a set of edges $E$. Let $q_{ik}$ be the flow in the edge $(i,k)$ and $%
Q_{i}$ be the supply or demand called intensity of node $i,$ which satisfies
the next conservation law:

\begin{equation}
\sum_{k}q_{ik}+Q_{i}=0,\quad  \label{d1}
\end{equation}

\begin{equation}
q_{ik}=-q_{ki},\quad i,k\in V,\quad (i,k)\in E.  \label{d1_ik}
\end{equation}

The minimal cost network flow problem consists in 
\begin{equation}
\text{minimize }F=\sum_{(i,k)\in E}F_{ik}(q_{ik})  \label{minCost}
\end{equation}%
\[
\text{subject to } 
\]

\begin{equation}
\sum_{k}q_{ik}-\sum_{m}q_{mi}=Q_{i}\text{,}\quad q_{ik}=-q_{ki},\quad i,k\in
V,\quad (i,k)\in E,  \label{qQCost}
\end{equation}%
\begin{equation}
q_{ik}^{-}\leq q_{ik}\leq q_{ik}^{+},\quad (i,k)\in E.  \label{qCost}
\end{equation}%
In the dual problem, the objective function depends on the differences of
potentials $p_{i},p_{k},$ i.e. on tension vectors $t_{ik}=p_{i}-p_{k}.$ The
necessary and sufficient optimality conditions for the minimum cost network
flow problems with linear or nonlinear objective functions are well known 
\cite{3 Himmelblau-72}, \cite{10 Murtagh-85}. . They include the connection
between pressures $p_{i},p_{k}$ and flow $q_{ik}$ in the form of equality
and inequality system between the pressure differences and the left and
right derivatives of the cost at the flow. They satisfy the following
complementary slackness condition \cite{EM-68}, \cite{ZM-86} in the
necessary and sufficient optimality conditions 
\begin{equation}
q_{ik}=q_{ik}^{-}\quad \text{if and only if }\quad p_{i}-p_{k}<\frac{%
\partial F_{ik}}{\partial q_{ik}}\mid q_{ik}^{-},  \label{qmi}
\end{equation}%
\begin{equation}
q_{ik}=q_{ik}^{+}\quad \text{if and only if }\quad p_{i}-p_{k}>\frac{%
\partial F_{ik}}{\partial q_{ik}}\mid q_{ik}^{+},  \label{qma}
\end{equation}%
\begin{equation}
q_{ik}=q_{ik}^{0}\quad \text{if and only if }\quad p_{i}-p_{k}=\frac{%
\partial F_{ik}}{\partial q_{ik}}\mid q_{ik}^{0}\text{ }\quad \text{for }%
\quad q_{ik}^{-}\leq q_{ik}^{0}\leq q_{ik}^{+}.  \label{q0}
\end{equation}%
While the edge $(i,k)$ represents the equipment with input node $i$ and
output node $k,$ the equality (\ref{q0}) 
\begin{equation}
p_{i}-p_{k}=\frac{\partial F_{ik}}{\partial q_{ik}}  \label{qrule}
\end{equation}%
represents the model of equipment, i.e. Ohm's law, Bernoulli's law etc.
Really, if $F_{ik}$ is linear,

\[
F_{ik}(q_{ik})=c_{ik}q_{ik}\quad \text{ then}\quad \text{ }%
p_i-p_k=c_{ik}\quad \text{ that corresponds cost problem,} 
\]
for quadratic $F_{ik}$

\[
F_{ik}(q_{ik})=c_{ik}q_{ik}^2\text{ }\quad \text{then }\quad p_i-p_k\sim
q_{ik}\quad \text{ that corresponds Ohm's law,} 
\]
for cubic $F_{ik}$

\[
F_{ik}(q_{ik})=c_{ik}q_{ik}^3\quad \text{ then}\quad \text{ }p_i-p_k\sim
q_{ik}^2\quad \text{ that corresponds Bernoulli's law.} 
\]

Let us generalize equipment models so that together with equations (\ref%
{qrule}) connecting potential difference with a function of flow there can
be considerable any function of flow and both potentials

\begin{equation}
f_{ik}(p_{i},p_{k},q_{ik})=0,\quad (i,k)\in E.  \label{equip}
\end{equation}%
To give a possibility to select the equipment, we come to the necessity to
select the model, i.e. to choice the best equation from the equation set, we
suppose that there are given the families of functional dependencies between
flow and pressures:

\begin{equation}
f_{d_{_{ik}}}(p_{i},p_{k},q_{ik},c_{ik})=0,\quad (i,k)\in E,  \label{d2}
\end{equation}

\begin{equation}
d_{ik}\in \{1,...,N_{ik}\}.  \label{d3}
\end{equation}

Here $c_{ik}$ is a vector of continuous parameters (coefficients), and $%
d_{ik}$ is a discrete parameter on the edge $(i,k)$. We suppose that there
are given the limitations $Q_{i}^{-},$ $Q_{i}^{+},$ $p_{i}^{-},$ $p_{i}^{+},$
$c_{ik}^{-},$ $c_{ik}^{+}$ :

\begin{equation}
Q_{i}^{-}\leq Q_{i}\leq Q_{i}^{+},\quad i\in V,  \label{d4}
\end{equation}

\begin{equation}
p_{i}^{-}\leq p_{i}\leq p_{i}^{+},\quad i\in V,  \label{d5}
\end{equation}

\begin{equation}
c_{ik}^{-}\leq c_{ik}\leq c_{ik}^{+},\quad (i,k)\in E,  \label{d6}
\end{equation}%
and the other restrictions can be represented by inequalities with given $%
a_{ik}^{-},$ $a_{ik}^{+}$ for the vector-functions $%
a_{ik}(p_{i},p_{k},q_{ik},c_{ik},d_{ik})$ which have to be calculated:

\begin{equation}
a_{ik}^{-}\leq a_{ik}(p_{i},p_{k},q_{ik},c_{ik},d_{ik})\leq a_{ik}^{+},\quad
(i,k)\in E.  \label{d7}
\end{equation}

Because the roles of input and output and the flow direction could be
changed, there is considered an undirected network.

The considered objective function depends both on flow qik and on pressures $%
p_{i},$ $p_{k},$ intensities $Q_{i},$ continuous $c_{ik}$ and discrete
parameters $d_{ik}$ . Then the problem is:

\begin{equation}
minimize\quad
F=\sum_{ik}F_{ik}^{(1)}(p_{i},p_{k},q_{ik},c_{ik},d_{ik})+%
\sum_{i}F_{i}^{(2)}(p_{i},Q_{i})  \label{d8}
\end{equation}

\begin{equation}
\text{subject to (\ref{d1}),(\ref{d1_ik}),(\ref{d2})-(\ref{d7}).}
\end{equation}

We may interpret (\ref{d7}) as restrictions on the power, temperature,
dissipation and other characteristics of the equipment that is represented
by the edge $(i,k)$. The set of available values of discrete parameters $%
d_{ik}$ in (\ref{d3}) means that the family of functions $f_{d_{ik}}$ can
act on the edge $(i,k)$, and we have to select the best function $f_{d_{ik}}$
for an equation (\ref{d2}). Thus, we have to select such equation (\ref{d2}%
), which is the best for the objective function (\ref{d8}). We may interpret
this as a selection of the most profitable equipment, which is installed or
can be installed on the place $(i,k)$.

The continuous parameters $c_{ik}$ in (\ref{d2}) can be interpreted as
parameters that smoothly regulate the work of equipment dik in bounds (\ref%
{d6}).

The inequalities (\ref{d4}) and the dependence of objective function on
intensities $Q_{i}$ mean that the most profitable values of supplies and
demands have to be chosen.

Without concentration on the conditions of existence and uniqueness of the
solution and on the proving of convergence to it, we describe the methods
that work in practice and bring the solution of the formulated problem in
the general case.

\subsection{Merits and demerits of different optimization problems}

\subsubsection{The network optimization based on hydraulic simulation.}

Advantages:

\qquad 1) The network optimization with hydraulic simulation could be useful
for both operational and economic optimization;

\qquad 2) The very broad range of technological restrictions can be
considered;

\qquad 3) The broad range of objective functions can be considered,
multi-objective problem can be modelled;

\qquad 4) Steady state network simulation is provided;

\qquad 5) Pressure is considered explicitly. Restrictions and objective
functions are formulated in terms of both pressure and flow;

\qquad 6) Exactly as in network operation, capacity of pipelines are
represented implicitly. This means that pipeline capacities are given due to
maximal and minimal pressure limits in nodes of network. Pressure limits
arise owing to technical and contractual restrictions;

\qquad 7) In comparison with optimization based only on flow models, the
network optimization based on hydraulic simulation has a high technical
reliability and exactness of results.

Demerits:

\qquad 1) Complication of models and methods;

\qquad 2) In comparison with optimization based only on flow models, the
network optimization based on hydraulic simulation usually needs more
computational time and more efforts of user to get final results.

\subsubsection{The network optimization based only on flow models.}

Merits:

\qquad 1) Simplicity of models;

\qquad 2) In comparison with network optimization based on hydraulic
simulation, the optimization based only on flow models usually needs not so
much computational time and less efforts of user to get final results

Demerits:

\qquad 1) Pressure is considered implicitly;

\qquad 2) Pipeline capacities must be given as input data. They simulate
technical and contractual restrictions such as maximal and minimal pressure
limits only indirectly;

\qquad 3) The optimization based only on flow models has rather low
technical reliability of results.

\section{Method of continuous - discrete nonlinear optimization on a network}

The formulated problem (\ref{d1}) - (\ref{d8}) is a problem of constrained
nonlinear mixed integer optimization, i.e. a problem of a search for an
extremum for a function, which has continuous - discrete parameters of
optimization and which undergoes nonlinear constraints. The objective
function, constraints and parameters of optimization here are given on a
network.. According to the mixing character of variables, the combination of
the continuous and discrete optimization methods is used to solve the
problem. The graph theory, integer programming, nonlinear programming with
and without constraints and the methods of solving the nonlinear systems of
equations and inequality are the basis of the offered method. Its main
characteristic feature consists in obtaining the dominant solutions on the
specially constructed subnetworks.

The discrete variable is e.g. the configuration of a compressor station,
i.e. how much compressors have to be in operation and what scheme of their
connection has to be used there. To manage the discrete variables, there are
at least two methods in ACCORD. The first one is a special modification of
branch-and-bound method. Another procedure for searching an integer-feasible
solution is connected with a possibility to check whether there is a
realization of equation $t_{jk}=p_{j}-p_{k}$ in the equation family (\ref{d2}%
). The complementary variable $t_{jk}$ will be penalized if it cannot be
represented by one of equations (\ref{d2}).

The difference between this method and the branch-and-bound approach is
clear for the cases of pure discrete problems on a tree. Such class of
problems means that the network has no cycles and there are no continuous
variables at all. Then the second method can fall in a local optimum. To
avoid it, a restart can be useful. For the same problem on a tree, the
branch- and-bound method can be very efficient. It brings a global optimum
as a solution. However, for a network with complicated topology the
branch-and-bound method could be expensive if not prohibited. This is valid
especially for the networks with a lot of cycles. Then the second procedure
consisting of searching the discrete variables $d_{jk}$ by penalizing
non-realizable continuous complementary variable $t_{jk}$ is available there.

In practice, the method avoids the local optima.

\section{Numerical example and experience: optimal operation of a gas network%
}

The graph of a tested gas network is presented on Fig. 1. The network
consists of a central distributed ring with large - scale supply and
transport pipelines. Every edge is either a pipeline section or a compressor
station. The input data for nodes are given in Table 1, for pipeline
sections in Table 2, for compressor stations in Table 3. The main benefits
of optimization are collected in Table 4.

As a first approximation for pressure drop on a pipeline, there is used the
simple formula for pressure drop on an aligned pipeline section

\begin{equation}
P_{i}^{2}-P_{e}^{2}=k_{D}lq_{ie}^{2}sign\left( q_{ie}\right) ,  \label{d9}
\end{equation}%
where: $P_{i}$ , $P_{e}$ - pressures respectively in inlet node $i$ and
outlet ("exit") node $e$, [bar]; $q_{ie}$ - flow rate [Mm%
${{}^3}$%
/d], i.e. gas flow volume per day reduced to the standard temperature $%
T=273K $ and pressure $P=1.013$ bar; $l$ - pipeline length, [km]; $k_{D}$ -
coefficient defined by formula

\begin{equation}
k_{D}=R_{a}d_{a}^{2}10^{-1}\lambda zGT/\left\{ E^{2}(\pi /4)\left( \sum
D_{i}^{5/2}\right) ^{2}\right\} ,  \label{d10}
\end{equation}%
where: $E$ - efficiency factor that is a relation between real and
theoretical flow such that $E<1$ and $E=0.92$ for average operating
conditions; $R_{a}=287$ J/kgK - gas constant for air, $d_{a}=1.293$ kg/m%
${{}^3}$
- air normal density; $\lambda $ - hydraulic resistance coefficient defined
e.g. after Colebrook - White formula \cite{11 Simone-88}, p. 286, $\left[ 1%
\right] $; $z$ - average gas compressibility, $\left[ 1\right] $; $G$ - gas
specific density on air (for air, $G_{air}=1$); $T$ - average gas
temperature, $[K]$; $D_{i}$ - inside diameter of the $i$-th parallel pipe, $%
[m]$; here $[1]$ means a dimensionless value.

.Finally, for pressure drop on (unaligned) pipelines, the more complicated
formula \cite{11 Simone-88} is used.

The simulation of a compressor station cannot be reduced to a formula like
it has been made for a pipeline section. However, as a result of the
simulating algorithm, the exhaust pressure $P_{e}$ can be computed as a
function of inlet pressure $P_{i}$ , flow rate $q_{ie}$ , gas
characteristics, and operating equipment of compressor station $a_{ie}$ and $%
d_{ie}$ :

\begin{equation}
p_{e}=f_{ei}(P_{i},q_{ie},T_{i},...,a_{ie},d_{ie}),  \label{d11}
\end{equation}%
where $a_{ie}$ are continuous characteristics of equipment and $d_{ie}$ are
discrete characteristics, i.e. compressor types and connection schemes of
compressors.

For our example, two problems have been solved:

\qquad 1) Reproduction of initial operating conditions by means of
optimization with criterion of minimal deviation from set points;

\qquad 2) Minimization of expenditure in the power expression.

It was required to support the set points of pressure in nodes of pipeline
joints. For this purpose, the narrow intervals of feasible pressure were
taken in these nodes. These intervals are presented in Table 1. The
intervals in other nodes are equal (0; 56 bar). The node 48 is taken as a
root with the root pressure $P_{48}=55$ bar.

The value $E=0.9$ of efficiency factor has been used in formula (\ref{d10})
for most of the pipeline sections. Efficiencies of other pipeline sections
were varied to reach the feasible pressure area and to reproduce the initial
operating conditions (column "$k_{D}$ initial" of Table 2). On compressor
stations, the reproduction of initial operating conditions has been provided
by using efficiency factor $E$ and with schemes of connection of equipment
which are presented in Table 3 in the columns "Efficiency factor" and
"Initial schemes" respectively. For compressor, efficiency $E$ is a relation
between real and theoretical productivity like it is for pipeline section.
With the help of it, the real characteristics of compressors are taken into
account.

In a lot of practical implementations the simplified sub-optimal way can be
sufficient. It consists of dividing the initial discrete-continuous problem
to the pure discrete problem on a spanning tree with computed flow and to
the prefix and postfix network continuous problems which could be solved
interactively. However, we solved the full optimization problem for the test
network.

In the central distributed ring two chords have been distinguished: the
pipeline sections 29 = (14, 17) and 35 = (22, 24). The pressure errors in
nodes 17 and 24 are corresponding to flows in these chords. The found flows
and the pressure errors are:

\qquad $q_{14,17}=22.654$ Mm%
${{}^3}$%
/d, \qquad \qquad $dP_{17}=0.01$ bar.

\qquad $q_{24,22}=5.906$ Mm%
${{}^3}$%
/d,$\qquad \qquad dP_{24}=0.33$ bar.

The optimal connection of equipment on compressor stations has been found.
The results are presented in Table 3 in the column "Optimal schemes". The
power expenditure is reduced from 260.765 MW to 234.432 MW that means 11.2
\% decreasing.

From the presented data we can see that the solution of the optimization
problem enables to find the feasible operating network conditions and to
reduce the transport expenditure by 11.2 \% as a result of selection of
optimal equipment on the compressor stations No. 1, 6, 12 and 18.

\section{Quality, supply, and cost tracking}

We present the quality, supply and cost tracking computed by ACCORD in
consequence of optimization (Fig. 2 - 9). The graphic presentation of the
tracking is made on a path between two nodes in a pipeline network. To
choice a path, user clicks on two end-nodes - on the origin and terminus of
the path. If the path between these nodes is unique, it will be selected by
ACCORD. Otherwise it should be necessary to click on one or a few additional
nodes.

Figures (i) 2 and 3, (ii) 4 and 5, (iii) 6 and 7, (iv) 8 and 9 can be
considered as four pairs of diagrams. In each of these pairs, the first
diagram shows the parameters distributed along the path: In Fig. 2, there
are distance from origin in km as axis x, pressure as axis y1 , and flow as
axis y2 . In the second diagram, i.e. here in Fig. 3, the parameters
concerning nodes are shown: There are supplies and demands situated on the
selected path and flows located in the pipelines branched off the path. In
Fig. 4, supply tracking is shown, while quality tracking as gas calorific
value is shown in Fig. 5.

The computed data presented in Figs. 2 and 3 and then in Figs. 4 and 5 are
basis for the supply cost tracking presented in Figs. 6 and 7. We can see
how the money for purchased gas is flowing along the pipeline per supply and
in total, and how much the normalized costs of purchased gas and the
addendum per supply for every consumer are. The operating expenses are
depending on the fuel costs. The flows of operating expenses and fix
(investment) cost are shown in Fig. 8. As a result, the normalized costs per
consumer are presented in Fig. 9.

\section{Conclusions}

The optimization methods and program ACCORD are developed for network
optimization. There are used steady state models of equipment. The equipment
such as pipelines, compressor stations, controlled valves, valves, etc. are
represented by the edges of the network. Supplies and demands, inlets and
outlets of equipment are represented by the nodes of the network. The
economic and operational problems solved in ACCORD are formulated as
continuous, discrete and discrete-continuous optimization on general network
with given family of functional connections between flow and pressure. The
continuous variables can be acted both in nodes and edges. Pressures, flow
and parameters of simulating models are examples of them. The discrete
parameters correspond to edges. Examples of them are internal structure of
edges and types of simulating models of edges, i. e. of concrete functional
dependencies between flow and pressure. The optimal equipment to operate can
be selected. The generalization of the discrete characteristic of nodes such
as "the node is a source (a sink)" is possible.

During optimization, the methods adapt themselves to the concrete network
and the features of both object function and variables.

There is proposed a model of contracts having a set of feasible intervals of
demand and pressure with corresponding procedures for payment calculation.
Every contract is represented by a (sub-) network. Such a model causes
different merits. It enables a simple, visualizable, structural and unified
representation of contracts.

For short term planning, the steady state optimization of gas network is a
useful tool. For long term planning for serial time periods, it is
worthwhile to use the minimal cost network flow problem as a compromise
simplification. The correct investigation of capacities can turn out as a
weakest point in such a simplified approach. Here a method \cite{9 0-93} is
useful; which we developed for precise definition of pressure margin in gas
networks.

Steady state optimization by ACCORD has been connected with dynamic
simulation programs and has been integrated in the distributed control
system GAMOS that consists of SCADA and various high level functions. ACCORD
possesses a clear and simple interface in the form of text files. Due to
this interface, ACCORD can be connected with any other simulation program,
distributed control system and SCADA system to provide steady state
optimization of gas networks with economic, business or operational
objectives.


\bigskip

\newpage

\section{Notation\textbf{\ }}

\textbf{Letters}

$A=$ area or cross section, $[m^{2}]$

$A_{k}=\frac{k-1}{k}$, $[-]$

$A_{n}=\frac{n-1}{n}$, $[-]$

$a=$ speed of sound, $[m/s]$

$b=$ outlet width of impeller, $[m]$

$C=$ conversion factor , $[-]$

$D=$ outlet diameter of impeller, $[m]$

$E=$ efficiency, $[-]$

$H=$ mass-specific enthalpy, $[kJ/kg]$

$H_{p}=$ specific polytropic work of compression i.e. specific polytropic
head, $[kJ/kg]$

$k=$ isentropic exponent, $[-]$

$m^{\prime }=$ mass flow, $[kg/s]$

$Ma=$ Mach number, $[-]$

$Mix=$ gas composition as a list of components, $[\%]$

$n=$ polytropic exponent i.e. polytropic volume exponent, $[-]$

$N=$ speed of rotation, $[1/s]$ (sometimes $[rps/rpm])$

$P=$ absolute pressure, $[bar]$

$R,R_{g}=$ specific gas constant, $[J/(kg\cdot K)]$

$R_{0}=$ universal gas constant, $[J/(kmol\cdot K)]$

$Re=$ Reynolds number, $[-]$

$S=$ pressure ratio, $[-]$

$T=$ absolute temperature, $[K]$

$u=$ velocity; linear tip speed, referred to $D$ of impeller, $[m/s]$

$v=$ velocity, $[m/s]$

$V=$ volume, $[m^{3}]$

$V^{\prime }=\dot{V}=$ volume flow, $[m^{3}/s]$

$W=$ power, $[W]$

$Y=$ specific work of compression i.e. specific head, $[kJ/kg]$

$Z=$ compressibility factor, $[-]$

$\mu =$ molar mass, $[kg/kmol]$

$\nu =$ kinematic viscosity, $[m^{2}/s]$

$\rho =$ density, $[kg/m^{3}]$

$\varphi =$ flow coefficient, $[-]$

$\psi =$ head coefficient i.e. process work coefficient, $[-]$

\bigskip

\textbf{Indexes}

$1=$ inlet

$2=$ outlet

$a=$ actual i.e. operating point of compressor

$e=$ expected value read from manufacturer characteristics

$c$ - converted point i.e. result of conversion of operating point of
compressor to the reference gas and the reference inlet conditions

$n=$ standard state i.e. standard condition for temperature and pressure
(STP)

$p=$ polytropic

$r$ - the reference gas composition and the reference state i.e. the
reference condition for temperature and pressure ($RTP$)

\bigskip


\bigskip

\newpage

\section{Figures}





\begin{figure}[tbp]
\hspace{10cm} \centering
\includegraphics[scale=0.95]{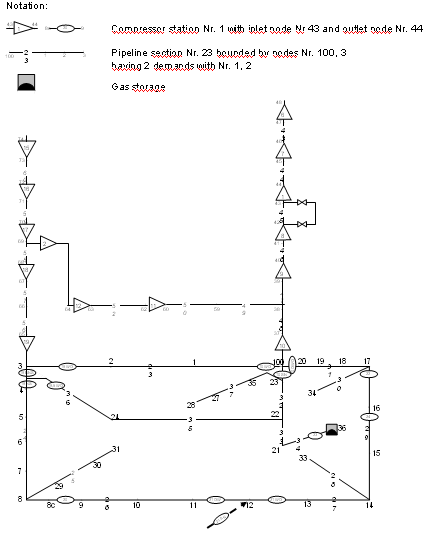}\newline
\caption{A gas pipeline network}
\label{fig:1}
\end{figure}
\bigskip

\bigskip

\begin{figure}[tbp]
\hspace{10cm} \centering
\includegraphics[scale=0.7]{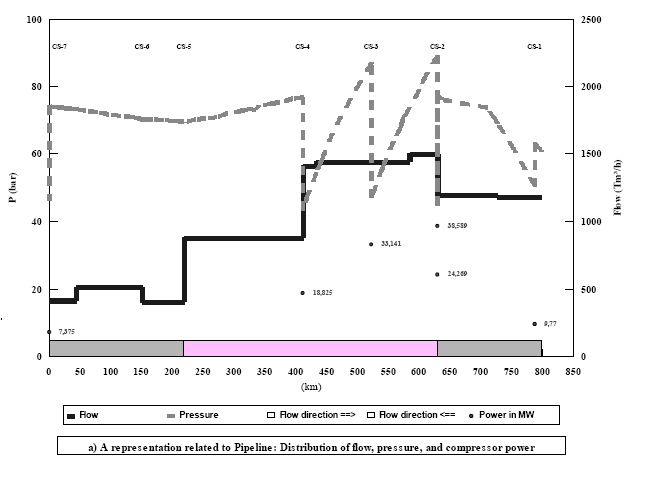}\newline
\caption{Pressure, flow, and the compressor power over the selected path of
the considered gas network }
\label{fig:2a}
\end{figure}
\bigskip

\bigskip

\begin{figure}[tbp]
\hspace{10cm} \centering\includegraphics[scale=0.7]{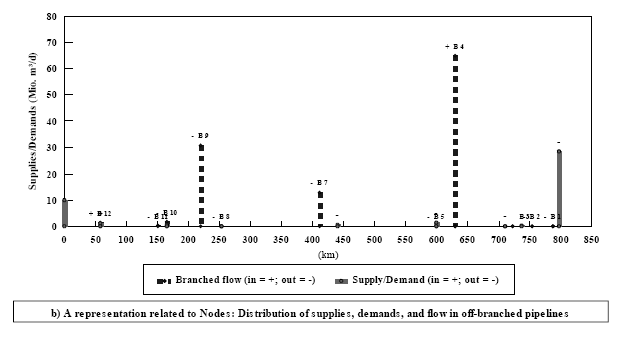}\newline
\caption{Supplies, demands, and flow in the off-branched egdes over the
selected path of the considered gas network }
\label{fig:2b}
\end{figure}

\bigskip \bigskip

\begin{figure}[tbp]
\hspace{10cm} \centering\includegraphics[scale=0.7]{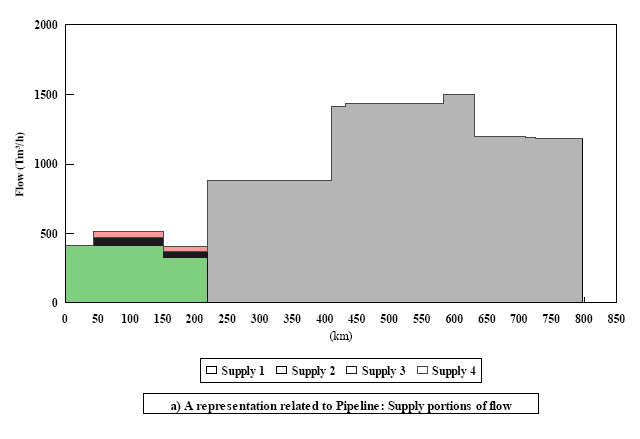}\newline
\caption{Supply portions in Tracking of supplies, gas composition,
thermodynamic properties, and quality parameters of the flow over the
selected path of the considered gas network}
\label{fig:3a}
\end{figure}

\bigskip

\bigskip

\begin{figure}[tbp]
\hspace{10cm} \centering\includegraphics[scale=0.7]{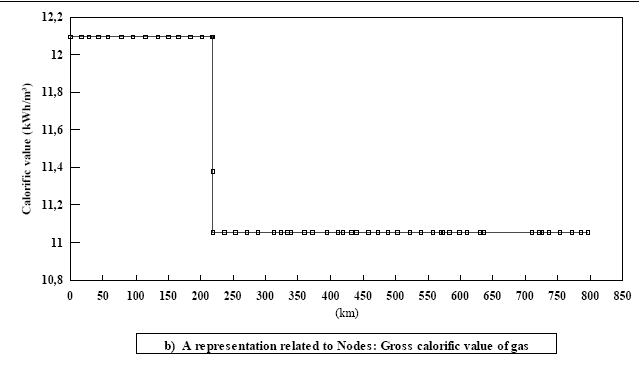}\newline
\caption{Heating value in Tracking of supplies, gas composition,
thermodynamic properties, and quality parameters of the flow over the
selected path of the considered gas network}
\label{fig:3b}
\end{figure}

\bigskip 
\begin{figure}[tbp]
\hspace{10cm} \centering\includegraphics[scale=0.7]{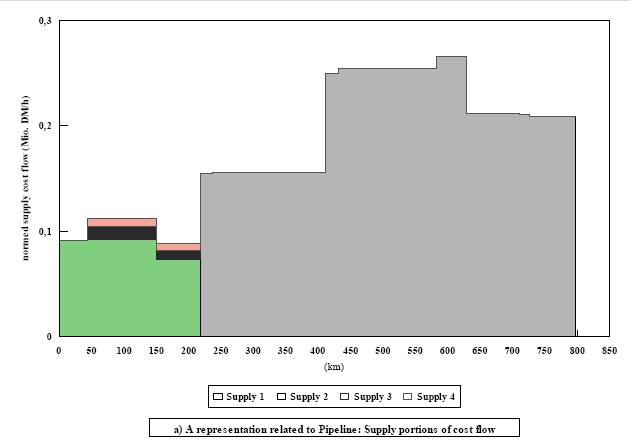}%
\newline
\caption{Costs of supplies in costs of flow as a part of Tracking of
supplies, gas composition, thermodynamic properties, and quality parameters
of the flow over the selected path of the considered gas network}
\label{fig:4a}
\end{figure}

\bigskip

\begin{figure}[tbp]
\hspace{10cm} \centering\includegraphics[scale=0.7]{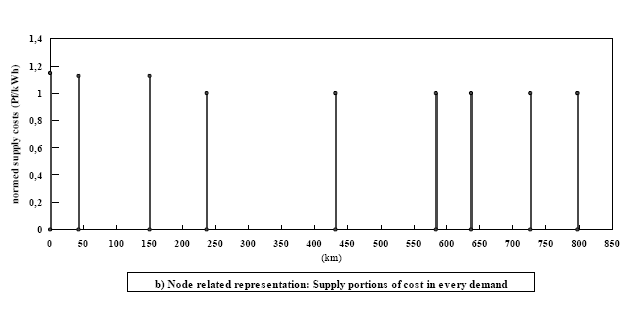}%
\newline
\caption{Tracking of the normed operating expenses: Costs of supplies in
costs of demands as a part of Tracking of supplies, gas composition,
thermodynamic properties, and quality parameters of the flow over the
selected path of the considered gas network}
\label{fig:4b}
\end{figure}

\bigskip

\begin{figure}[tbp]
\hspace{10cm} \centering\includegraphics[scale=0.7]{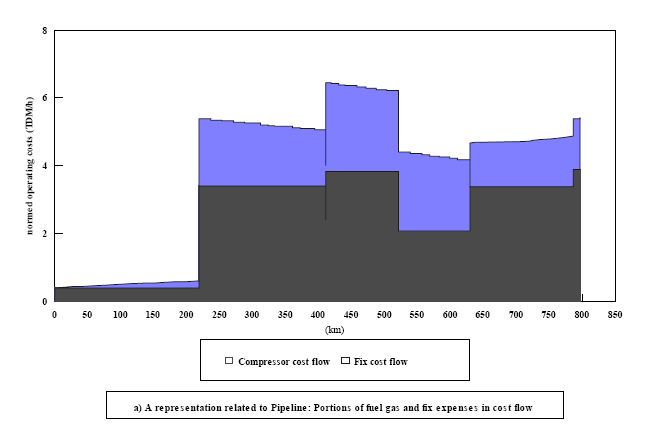}%
\newline
\caption{Costs of fuel gas, electro energy, and fix expenses in costs of
flow as a part of Tracking of supplies, gas composition, thermodynamic
properties, and quality parameters of the flow over the selected path of the
considered gas network}
\label{fig:5a}
\end{figure}

\bigskip

\begin{figure}[tbp]
\hspace{10cm} \centering\includegraphics[scale=0.65]{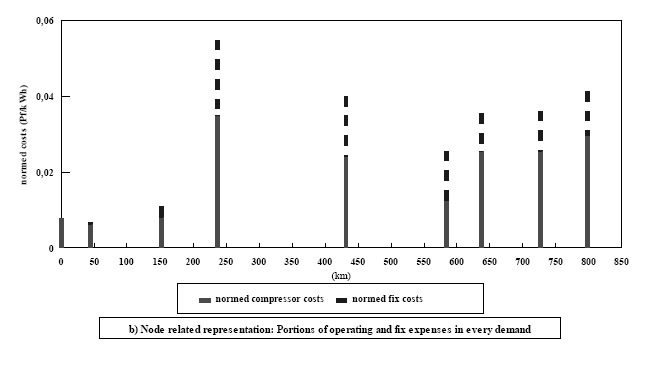}%
\newline
\caption{Transport and fix expenses: Costs of fuel gas, electro energy, and
fix expenses in costs of demands as a part of Tracking of supplies, gas
composition, thermodynamic properties, and quality parameters of the flow
over the selected path of the considered gas network}
\label{fig:5b}
\end{figure}

\bigskip


\bigskip

\newpage

\bigskip

\section{Tables}

\bigskip

Table 1. Sub-table 1.1. Data in end-nodes of the gas network

\bigskip

\begin{tabular}{lllllll}
\textbf{Cons.} & \textbf{Node} & \textbf{+, if } & \textbf{Supply or } & 
\textbf{P} & \textbf{Pmin} & \textbf{Pmax} \\ 
\textbf{No} &  & \textbf{supply} & \textbf{Demand,} &  &  &  \\ 
&  &  &  &  &  &  \\ 
&  &  & \textbf{Mm3/d} & \textbf{bar} & \textbf{bar} & \textbf{bar} \\ 
&  &  &  &  &  &  \\ 
1 & \textbf{100} &  & 2.3 & 41 & 40 & 42 \\ 
2 & \textbf{3} &  & 3.55 & 46 & 45 & 47 \\ 
3 & \textbf{8} &  & 1.9 &  &  & 56 \\ 
4 & \textbf{12} & + & 43.4 &  &  & 56 \\ 
5 & \textbf{21} &  & 9.83 &  &  & 56 \\ 
6 & \textbf{22} &  & 46.07 &  &  & 56 \\ 
7 & \textbf{24} &  & 8.14 &  &  & 56 \\ 
8 & \textbf{28} &  & 12.06 &  &  & 56 \\ 
9 & \textbf{31} &  &  &  &  & 56 \\ 
10 & \textbf{33} &  &  &  &  & 56 \\ 
11 & \textbf{34} &  & 10.76 &  &  & 56 \\ 
12 & \textbf{36} & + & 13.5 &  &  & 56 \\ 
13 & \textbf{37} &  &  & 41 & 40 & 42 \\ 
14 & \textbf{38} &  &  & 40 & 38 & 40.5 \\ 
15 & \textbf{39} &  &  & 28.3 & 27 & 29 \\ 
16 & \textbf{40} &  & 11.3 & 33.4 & 29 & 34 \\ 
17 & \textbf{41} & + & 19.9 & 31.3 &  & 56 \\ 
18 & \textbf{42} &  & 1.6 & 42.5 &  & 56 \\ 
19 & \textbf{43} &  & 0.1 & 37.5 &  & 56 \\ 
20 & \textbf{44} &  & 10.1 & 37.5 &  & 56%
\end{tabular}

\bigskip


\bigskip

\newpage

\bigskip

Table 1. Sub-table 1.2. Data in end-nodes of the gas network

\bigskip

\begin{tabular}{lllllll}
\textbf{Cons.} & \textbf{Node} & \textbf{+, if} & \textbf{Supply or } & 
\textbf{P} & \textbf{Pmin} & \textbf{Pmax} \\ 
\textbf{No.} &  & \textbf{supply} & \textbf{Demand,} &  &  &  \\ 
&  &  &  &  &  &  \\ 
&  &  & \textbf{Mm3/d} & \textbf{bar} & \textbf{bar} & \textbf{bar} \\ 
&  &  &  &  &  &  \\ 
21 & \textbf{45} &  & 0.1 & 33.5 &  & 56 \\ 
22 & \textbf{46} &  & 7.4 & 41.1 &  & 56 \\ 
23 & \textbf{47} & + & 65.2 & 37.4 &  & 56 \\ 
24 & \textbf{48} &  & 77.4 & 55 &  & 56 \\ 
25 & \textbf{59} &  & 16.3 & 43 &  & 56 \\ 
&  &  &  &  &  &  \\ 
26 & \textbf{60} &  & 5.7 & 47.7 &  & 56 \\ 
27 & \textbf{62} &  &  & 33.7 &  & 56 \\ 
28 & \textbf{63} &  &  & 49.6 &  & 56 \\ 
29 & \textbf{64} & + & 35.4 & 38.4 &  & 56 \\ 
30 & \textbf{65} &  &  & 33.9 &  & 56 \\ 
31 & \textbf{67} &  & 0.4 & 51 &  & 56 \\ 
32 & \textbf{68} &  & 0.4 & 34.3 &  & 56 \\ 
33 & \textbf{69} &  & 0.2 & 53.4 &  & 56 \\ 
34 & \textbf{70} &  & 2.6 & 36.4 &  & 56 \\ 
35 & \textbf{71} &  & 0.4 & 52.5 &  & 56 \\ 
36 & \textbf{72} &  & 0.2 & 35.2 &  & 56 \\ 
37 & \textbf{73} &  & 0.2 & 52.5 &  & 56 \\ 
38 & \textbf{74} & + & 84.5 & 41 & 30 & 40%
\end{tabular}

\bigskip


\bigskip

\newpage

\bigskip

Table 2. Sub-table 2.1. Data of pipeline sections of the gas network

\bigskip

\begin{tabular}{lllllllll}
1 & 2 & 3 & 4 & 5 & 6 & 7 & 8 & 9 \\ 
No & Sec- & End- & L & Inter-. & Inter- & Ko & Ko & $\Delta $Ko \\ 
& tion & nodes & Sub- & medi- & medi- & initial & result &  \\ 
&  &  & sec- & ate & ate &  &  &  \\ 
&  &  & tion & Node & Demand &  &  &  \\ 
&  &  &  &  &  &  &  &  \\ 
& -- & -- & km & -- & Mm3/d & -- & -- & \% \\ 
&  &  &  &  &  &  &  &  \\ 
\textbf{1} & \textbf{23} & 100 -- 3 & 36 & 1 & 0.5 & 0.083258 & 0.005263 & 
-93.7\% \\ 
&  &  & 25 & 2 & 2.5 &  &  &  \\ 
&  &  & 38 &  &  &  &  &  \\ 
&  &  &  &  &  &  &  &  \\ 
\textbf{2} & \textbf{24} & 3 -- 8 & 28 & 4 & 0.6 & 0.011032 & 0.011032 & --
\\ 
&  &  & 23 & 5 & 1 &  &  &  \\ 
&  &  & 19 & 6 & 2 &  &  &  \\ 
&  &  & 16 & 7 & 3.9 &  &  &  \\ 
&  &  & 21 &  &  &  &  &  \\ 
&  &  &  &  &  &  &  &  \\ 
\textbf{3} & \textbf{25} & 8 -- 31 & 9 & 29 & 1.2 & 0.011032 & 0.011032 & --
\\ 
&  &  & 14 & 30 &  &  &  &  \\ 
&  &  & 8 &  &  &  &  &  \\ 
&  &  &  &  &  &  &  &  \\ 
\textbf{4} & \textbf{26} & 8 -- 12 & 28 & 9 & 2.7 & 0.83258 & 0.715182 & 
-14.1\% \\ 
&  &  & 13 & 10 & 1 &  &  &  \\ 
&  &  & 31 & 11 & 1.2 &  &  &  \\ 
&  &  & 1 &  &  &  &  &  \\ 
&  &  &  &  &  &  &  &  \\ 
\textbf{5} & \textbf{27} & 12 -- 14 & 9 & 13 & 0.3 & 0.005899 & 0.005899 & --
\\ 
&  &  & 38 &  &  &  &  &  \\ 
&  &  &  &  &  &  &  & 
\end{tabular}

\bigskip


\bigskip

\newpage

\bigskip

Table 2. Sub-table 2.2. Data of pipeline sections of the gas network

\bigskip

\begin{tabular}{lllllllll}
1 & 2 & 3 & 4 & 5 & 6 & 7 & 8 & 9 \\ 
No & Sec- & End- & L & Inter-. & Inter- & Ko & Ko & $\Delta $Ko \\ 
& tion & nodes & Sub- & medi- & medi- & initial & result &  \\ 
&  &  & sec- & ate & ate &  &  &  \\ 
&  &  & tion & Node & Demand &  &  &  \\ 
&  &  &  &  &  &  &  &  \\ 
& -- & -- & km & -- & Mm3/d & -- & -- & \% \\ 
&  &  &  &  &  &  &  &  \\ 
\textbf{6} & \textbf{28} & 14 -- 33 & 28 & 32 & 2.1 & 0.083258 & 0.083258 & 
-- \\ 
&  &  & 5 &  &  &  &  &  \\ 
&  &  &  &  &  &  &  &  \\ 
\textbf{7} & \textbf{29} & 14 -- 17 & 9 & 15 & 3.8 & 0.005899 & 0.005899 & --
\\ 
&  &  & 15 & 16 & 1.4 &  &  &  \\ 
&  &  & 18 &  &  &  &  &  \\ 
&  &  &  &  &  &  &  &  \\ 
\textbf{8} & \textbf{30} & 17 -- 34 & 16 &  &  & 0.16232 & 0.16232 & -- \\ 
&  &  &  &  &  &  &  &  \\ 
\textbf{9} & \textbf{31} & 17 -- 100 & 7 & 18 & 0.07 & 0.083258 & 0.083258 & 
-- \\ 
&  &  & 22 & 19 & 0.07 &  &  &  \\ 
&  &  & 35 & 20 & 0.05 &  &  &  \\ 
&  &  & 38 &  &  &  &  &  \\ 
&  &  &  &  &  &  &  &  \\ 
\textbf{10} & \textbf{32} & 100 -- 22 & 68 & 23 & 4.3 & 0.792462 & 0.792462
& -- \\ 
&  &  & 11 &  &  &  &  &  \\ 
&  &  &  &  &  &  &  & 
\end{tabular}

\bigskip


\bigskip

\newpage

\bigskip

Table 2. Sub-table 2.3. Data of pipeline sections of the gas network

\bigskip

\begin{tabular}{lllllllll}
1 & 2 & 3 & 4 & 5 & 6 & 7 & 8 & 9 \\ 
No & Sec- & End- & L & Inter-. & Inter- & Ko & Ko & $\Delta $Ko \\ 
& tion & nodes & Sub- & medi- & medi- & initial & result &  \\ 
&  &  & sec- & ate & ate &  &  &  \\ 
&  &  & tion & Node & Demand &  &  &  \\ 
&  &  &  &  &  &  &  &  \\ 
& -- & -- & km & -- & Mm3/d & -- & -- & \% \\ 
&  &  &  &  &  &  &  &  \\ 
\textbf{11} & \textbf{33} & 22 -- 21 & 6 &  &  & 0.792462 & 0.792462 & -- \\ 
&  &  &  &  &  &  &  &  \\ 
\textbf{12} & \textbf{34} & 21 -- 36 & 75 &  &  & 0.872981 & 0.872981 & --
\\ 
&  &  &  &  &  &  &  &  \\ 
\textbf{13} & \textbf{35} & 22 -- 24 & 12 &  &  & 0.16232 & 0.16232 & -- \\ 
&  &  &  &  &  &  &  &  \\ 
\textbf{14} & \textbf{36} & 24 -- 3 & 106.3 &  &  & 0.027284 & 0.027284 & --
\\ 
&  &  &  &  &  &  &  &  \\ 
\textbf{15} & \textbf{37} & 100 -- 28 & 13 & 35 & 1.2 & 0.083258 & 0.083258
& -- \\ 
&  &  & 80 & 27 & 3 &  &  &  \\ 
&  &  &  &  &  &  &  &  \\ 
\textbf{16} & \textbf{43} & 47 -- 46 & 12 &  &  &  & 0.00673 &  \\ 
&  &  &  &  &  &  &  &  \\ 
\textbf{17} & \textbf{44} & 44 -- 45 & 100 &  &  & 0.00673 & 0.00673 & -- \\ 
&  &  &  &  &  &  &  &  \\ 
\textbf{18} & \textbf{45} & 42 -- 43 & 100 &  &  & 0.311 & 0.311 & -- \\ 
&  &  &  &  &  &  &  &  \\ 
\textbf{19} & \textbf{46} & 40 -- 41 & 100 &  &  & 0.0102 & 0.011 & 7.8\% \\ 
&  &  &  &  &  &  &  &  \\ 
\textbf{20} & \textbf{47} & 38 -- 39 & 100 &  &  & 0.00448 & 0.0166 & 270.5\%
\\ 
&  &  &  &  &  &  &  & 
\end{tabular}

\bigskip


\bigskip

\newpage

\bigskip

Table 2. Sub-table 2.4. Data of pipeline sections of the gas network

\bigskip

\begin{tabular}{lllllllll}
1 & 2 & 3 & 4 & 5 & 6 & 7 & 8 & 9 \\ 
No & Sec- & End- & L & Inter-. & Inter- & Ko & Ko & $\Delta $Ko \\ 
& tion & nodes & Sub- & medi- & medi- & initial & result &  \\ 
&  &  & sec- & ate & ate &  &  &  \\ 
&  &  & tion & Node & Demand &  &  &  \\ 
&  &  &  &  &  &  &  &  \\ 
& -- & -- & km & -- & Mm3/d & -- & -- & \% \\ 
&  &  &  &  &  &  &  &  \\ 
\textbf{21} & \textbf{48} & 38 -- 37 & 100 &  &  & 0.00918 & 0.00262 & 
-71.5\% \\ 
&  &  &  &  &  &  &  &  \\ 
\textbf{22} & \textbf{49} & 59 -- 38 & 100 &  &  & 0.00282 & 0.00282 & -- \\ 
&  &  &  &  &  &  &  &  \\ 
\textbf{23} & \textbf{50} & 60 -- 59 & 100 &  &  & 0.00483 & 0.00483 & -- \\ 
&  &  &  &  &  &  &  &  \\ 
\textbf{24} & \textbf{52} & 63 -- 62 & 100 &  &  & 0.00786 & 0.00786 & -- \\ 
&  &  &  &  &  &  &  &  \\ 
\textbf{25} & \textbf{56} & 66 -- 65 & 100 &  &  & 0.00037 & 0.00037 & -- \\ 
&  &  &  &  &  &  &  &  \\ 
\textbf{26} & \textbf{57} & 67 -- 66 & 100 &  &  & 0.00189 & 0.00189 & -- \\ 
&  &  &  &  &  &  &  &  \\ 
\textbf{27} & \textbf{58} & 69 -- 68 & 100 &  &  & 0.00255 & 0.00255 & -- \\ 
&  &  &  &  &  &  &  &  \\ 
\textbf{28} & \textbf{59} & 71 -- 70 & 100 &  &  & 0.00204 & 0.00204 & -- \\ 
&  &  &  &  &  &  &  &  \\ 
\textbf{29} & \textbf{60} & 73 -- 72 & 100 &  &  & 0.0213 & 0.0213 & -- \\ 
&  &  &  &  &  &  &  &  \\ 
\textbf{30} & \textbf{86} & 61 -- 64 & 462.08 &  &  & 0.01636 & 0.01636 & --
\\ 
&  &  &  &  &  &  &  & 
\end{tabular}

\bigskip


\bigskip

\newpage

\bigskip

Table 3. Sub-table 3.1. Data of compressor stations belonging to the
sub-system No. 1 of the considered gas network

\bigskip

\begin{tabular}{llllllllllll}
1 & 2 & 3 & 4 & 5 & 6 & 7 & 8 & 9 & 10 & 11 & 12 \\ 
No & Com- & Effi- & T & Shops & Type & Po- & Num- & Ma- & Ma- & Change & Be-
\\ 
& pres- & cien- & gas & Num- &  & wer & ber & chine & chine &  & ne- \\ 
& sor & cy & inlet & ber & of & of & of & Con- & Con- &  & fit, \\ 
&  & of &  &  & Ma- & Com- & Ma- & figu- & figu- &  & $\Delta $ \\ 
& Sta- & Sta- &  &  & chi- & pres- & chi- & ra- & ra- &  & ma- \\ 
& tion & tion &  &  & nes & sors & nes & tion & tion &  & chi- \\ 
& No &  &  &  & Dri- &  & To- & Ini- & Opti- &  & nes \\ 
&  &  &  &  & vers &  & tal & tial & mal &  &  \\ 
&  &  &  &  &  &  &  &  &  &  &  \\ 
&  &  & C &  &  & MW &  &  &  &  &  \\ 
&  &  &  &  &  &  &  &  &  &  &  \\ 
\textbf{1} & \textbf{6} & 0.69 & 13 & 3 & turbine & 4 & 8 & 3 x 2 & 3 x 1 & 
changed & $-3$ \\ 
&  &  &  &  & turbine & 6 & 3 & 1 x 2 & 1 x 1 & changed & $-1$ \\ 
&  &  &  &  & electro & 4 & 8 & 3 x 2 & 2 x 1 & changed & $-4$ \\ 
&  &  &  &  &  &  &  &  &  &  &  \\ 
\textbf{2} & \textbf{7} & 0.91 & 9 & 3 & turbine & 4 & 5 & 2 x 1 & 2 x 1 & 
&  \\ 
&  &  &  &  & electro & 4 & 8 &  &  &  &  \\ 
&  &  &  &  & turbine & 5 & 3 &  &  &  &  \\ 
&  &  &  &  &  &  &  &  &  &  &  \\ 
\textbf{3} & \textbf{1} & 0.91 & 8 & 1 & electro & 4 & 16 & -- & 2 x 1 & 
changed & $+2$ \\ 
&  &  &  &  &  &  &  &  &  &  &  \\ 
\textbf{4} & \textbf{8} & 0.91 & 11 & 2 & electro & 4 & 8 & 1 x 2 & 1 x 2 & 
&  \\ 
&  &  &  &  & turbine & 4 & 5 & 2 x 2 & 2 x 2 &  &  \\ 
&  &  &  &  &  &  &  &  &  &  &  \\ 
\textbf{5} & \textbf{9} & 0.91 & 2 & 2 & electro & 4 & 7 & 2 x 1 & -- & 
changed & $-2$ \\ 
&  &  &  &  & electro & 4.5 & 5 &  &  &  &  \\ 
&  &  &  &  &  &  &  &  &  &  &  \\ 
\textbf{6} & \textbf{10} & 0.91 & 3 & 1 & electro & 4 & 8 &  &  &  &  \\ 
&  &  &  &  &  &  &  &  &  &  & 
\end{tabular}

\bigskip


\bigskip

\newpage

\bigskip

Table 3. Sub-table 3.2. Data of compressor stations belonging to the
sub-system No. 2 of the considered gas network

\bigskip

\begin{tabular}{llllllllllll}
1 & 2 & 3 & 4 & 5 & 6 & 7 & 8 & 9 & 10 & 11 & 12 \\ 
No & Com- & Effi- & T & Shops & Type & Po- & Num- & Ma- & Ma- & Change & Be-
\\ 
& pres- & cien- & gas & Num- &  & wer & ber & chine & chine &  & ne- \\ 
& sor & cy & inlet & ber & of & of & of & Con- & Con- &  & fit, \\ 
&  & of &  &  & Ma- & Com- & Ma- & figu- & figu- &  & $\Delta $ \\ 
& Sta- & Sta- &  &  & chi- & pres- & chi- & ra- & ra- &  & ma- \\ 
& tion & tion &  &  & nes & sors & nes & tion & tion &  & chi- \\ 
& No &  &  &  & Dri- &  & To- & Ini- & Opti- &  & nes \\ 
&  &  &  &  & vers &  & tal & tial & mal &  &  \\ 
&  &  &  &  &  &  &  &  &  &  &  \\ 
&  &  & C &  &  & MW &  &  &  &  &  \\ 
&  &  &  &  &  &  &  &  &  &  &  \\ 
\textbf{7} & \textbf{11} & 0.98 & 13 & 1 & electro & 4 & 7 & 3 x 2 & 3 x 1 & 
changed & $-3$ \\ 
&  &  &  &  &  &  &  &  &  &  &  \\ 
\textbf{8} & \textbf{12} & 0.98 & 13 & 1 & electro & 4 & 7 & 3 x 1 & 3 x 1 & 
&  \\ 
&  &  &  &  &  &  &  &  &  &  & 
\end{tabular}

\bigskip


\bigskip

\newpage

\bigskip

Table 3. Sub-table 3.3. Data of compressor stations belonging to the
sub-system No. 3 of the considered gas network

\bigskip

\begin{tabular}{llllllllllll}
1 & 2 & 3 & 4 & 5 & 6 & 7 & 8 & 9 & 10 & 11 & 12 \\ 
No & Com- & Effi- & T & Shops & Type & Po- & Num- & Ma- & Ma- & Change & Be-
\\ 
& pres- & cien- & gas & Num- &  & wer & ber & chine & chine &  & ne- \\ 
& sor & cy & inlet & ber & of & of & of & Con- & Con- &  & fit, \\ 
&  & of &  &  & Ma- & Com- & Ma- & figu- & figu- &  & $\Delta $ \\ 
& Sta- & Sta- &  &  & chi- & pres- & chi- & ra- & ra- &  & ma- \\ 
& tion & tion &  &  & nes & sors & nes & tion & tion &  & chi- \\ 
& No &  &  &  & Dri- &  & To- & Ini- & Opti- &  & nes \\ 
&  &  &  &  & vers &  & tal & tial & mal &  &  \\ 
&  &  &  &  &  &  &  &  &  &  &  \\ 
&  &  & C &  &  & MW &  &  &  &  &  \\ 
&  &  &  &  &  &  &  &  &  &  &  \\ 
\textbf{9} & \textbf{15} & 0.91 & 16 & 2 & electro & 4 & 10 & 4 x 2 & 4 x 2
&  &  \\ 
&  &  &  &  & turbine & 10 & 3 & 1 x 2 & 1 x 2 &  &  \\ 
&  &  &  &  &  &  &  &  &  &  &  \\ 
\textbf{10} & \textbf{16} & 0.83 & 20 & 2 & turbine & 10 & 6 & 2 x 2 & 2 x 2
&  &  \\ 
&  &  &  &  & turbine & 6 & 6 & 2 x 2 & 2 x 2 &  &  \\ 
&  &  &  &  &  &  &  &  &  &  &  \\ 
\textbf{11} & \textbf{17} & 0.83 & 14 & 2 & turbine & 10 & 3 & 1 x 2 & 1 x 2
&  &  \\ 
&  &  &  &  & electro & 4 & 10 & 4 x 2 & 4 x 2 &  &  \\ 
&  &  &  &  &  &  &  &  &  &  &  \\ 
\textbf{12} & \textbf{18} & 0.8 & 9 & 2 & turbine & 6 & 6 & 2 x 2 & 3 x 2 & 
changed & $+2$ \\ 
&  &  &  &  & electro & 4 & 10 & 4 x 2 & 2 x 2 & changed & $-4$ \\ 
&  &  &  &  &  &  &  &  &  &  &  \\ 
\textbf{13} & \textbf{19} & 0.98 & 6 & 2 & turbine & 5 & 5 & 2 x 2 & 2 x 2 & 
&  \\ 
&  &  &  &  & turbine & 6 & 6 & 2 x 2 & 2 x 2 &  &  \\ 
&  &  &  &  &  &  &  &  &  &  & 
\end{tabular}

\bigskip


\bigskip

\newpage

\bigskip

Table 4. Benefits in the energy expended on compressor stations of the gas
network

\bigskip

\begin{tabular}{llll}
1 & 2 & 3 & 4 \\ 
Initial & Optimal & Savings & Benefits \\ 
Total & Total & in & $\Delta $machines \\ 
Compressor & Compressor & Compressor &  \\ 
Power & Power & Power &  \\ 
&  &  &  \\ 
MW & MW & \% & -- \\ 
&  &  &  \\ 
260.8 & 234.4 & -11.2\% & -13 \\ 
&  &  & 
\end{tabular}

\end{document}